\documentstyle[epsfig,aps,prd,twocolumn]{revtex}

\def\etal{{\it et al.}}

\def\ie{{\it i.e.}}

\def\lsim{\mathrel {\vcenter {\baselineskip 0pt \kern 0pt
    \hbox{$<$} \kern 0pt \hbox{$\sim$} }}}
    \def\gsim{\mathrel {\vcenter {\baselineskip 0pt \kern 0pt
    \hbox{$>$} \kern 0pt \hbox{$\sim$} }}}
\def\dddot#1{{\buildrel {. \kern-.05em . \kern-.05em .} \over {#1}}}

\begin{document}

\title{THE HAWKING-UNRUH TEMPERATURE\\
 AND DAMPING IN A LINEAR FOCUSING CHANNEL}

\author{KIRK T.~McDONALD}
\address{Joseph Henry Laboratories, Princeton University, Princeton, NJ 08544
\\
{\tt mcdonald@puphep.princeton.edu
\\
http://puhep1.princeton.edu/$\tilde{\phantom{a}}$mcdonald/accel/}
\\
(January 30, 1998)
}

\maketitle

\begin{abstract}
The Hawking-Unruh effective temperature, $\hbar a^\star \over 2 \pi ck$,  due to
quantum fluctuations in the radiation of an accelerated charged-particle beam
can be used to show that transverse oscillations of the beam in a
practical linear focusing channel damp to the quantum-mechanical limit.  
A comparison is made between this behavior and that of beams in a wiggler.
\end{abstract}

\section{Introduction}

Many of the effects of quantum fluctuations on the behavior of charged
particles can be summarized concisely by an effective temperature first
introduced in gravitational fields by Hawking \cite{Hawking}, and
applied to accelerated particles (with the neglect of gravity) by 
Unruh \cite{Unruh}.

Hawking argued that the effect of the strong gravitational field of a black
hole on the quantum fluctuations of the surrounding space is to cause 
the black hole to radiate with a temperature
\begin{equation}
T = {\hbar g \over 2\pi ck},
\label{eq1}
\end{equation}
where $g$ is the acceleration due to gravity at the surface of the black hole,
$c$ is the speed of light, and $k$ is Boltzmann's constant.  Shortly
thereafter, Unruh argued that an accelerated observer should become excited
by quantum fluctuations to a temperature
\begin{equation}
T = {\hbar a^\star \over 2\pi ck},
\label{eq2}
\end{equation}
where $a^\star$ is the acceleration of the observer in its instantaneous rest
frame.  

In a series of papers, Bell and co-workers \cite{Bell}, have noted that
electron storage rings provide a demonstration of the utility of the 
Hawking-Unruh temperature (\ref{eq2}), with emphasis on the question of the 
incomplete polarization of the electrons due to quantum fluctuations of
synchrotron radiation.  The author has commented on how the Hawking-Unruh
temperature can be used to characterize quickly the limits on damping of
the phase volume of beams in electron storage rings \cite{McDonald87}, leading 
to well-known results of Sands \cite{Sands}.

\section{Quantum Analysis of a Linear Focusing Channel}

Recently, Chen, Huang and Ruth have discussed radiation damping in a linear
focusing channel \cite{Huang1,Huang2,Huang3},
finding that in such devices the beam can be 
damped to the quantum mechanical limit set by the uncertainty principle.
I show here how this result follows very quickly from an application of
the Hawking-Unruh temperature.

A linear focusing channel is a beam-transport system that confines the
motion of a charged particle along a straight central ray via a potential that
is quadratic in the transverse spatial coordinates.   This potential can
be characterized by a spring constant $k$, and hence the frequency $\omega$
of transverse oscillations (as observed in the laboratory frame) of a
particle of mass $m$ and Lorentz factor $\gamma$ is
\begin{equation}
\omega = \sqrt{k \over \gamma m}.
\label{eq3}
\end{equation}
If the amplitude of the oscillation in transverse coordinate $x$ is called
$x_0$, then the amplitude $a_0$ of the corresponding transverse acceleration is
\begin{equation}
a_0 = x_0 \omega^2 = {k x_0 \over \gamma m}.
\label{eq4}
\end{equation}

To apply the Hawking-Unruh temperature, we consider the motion
in the instantaneous rest frame of the particle.  Supposing the transverse
oscillations are small, the instantaneous rest frame is very nearly the
frame in which the particle has no longitudinal motion.  Quantities measured
in the instantaneous rest frame will by denoted with the superscript $\star$.
Thus, in the instantaneous rest frame the amplitude of the transverse 
acceleration as measured is
\begin{equation}
a^\star_0 = \gamma^2 a_0 = {\gamma k x_0 \over m},
\label{eq5}
\end{equation}
the frequency of the oscillation is
\begin{equation}
\omega^\star = \gamma \omega,
\label{eq6}
\end{equation}
and hence the transverse spring constant of the focusing channel appears as
\begin{equation}
k^\star = m \omega^{\star 2} = \gamma k.
\label{eq7}
\end{equation}

In the instantaneous rest frame, the charge particle finds itself in a
bath of radiation of characteristic temperature given by eq.~(\ref{eq2})
with acceleration $a^\star$ given by eq.~(\ref{eq5}).  This bath can be
regarded as the effect of quantum fluctuations, which excite transverse
oscillations (having two degrees of freedom) to characteristic energy 
$U^\star$ (as measured in the instantaneous rest frame) given by
\begin{equation}
U^\star = kT = {\hbar a^\star_0 \over 2 \pi c} = {\hbar \gamma k x_0 \over 2 \pi
m c}.
\label{eq8}
\end{equation}
The energy of transverse oscillation can also be written in terms of
the (invariant) transverse amplitude $x_0$ as
\begin{equation}
U^\star = {k^\star x_0^{\star 2}\over 2} = {\gamma k x_0^2 \over 2}.
\label{eq9}
\end{equation}
Hence, the amplitude of excitation of the transverse oscillations is
\begin{equation}
x_0 = {\hbar \over \pi m c} = {\lambdabar_C \over \pi},
\label{eq10}
\end{equation}
where $\lambdabar_C$ is the (reduced) Compton wavelength of the particle.

The amplitude (\ref{eq10}) must, however, be compared to the amplitude
of the zero-point oscillations of the system, considered as a quantum
oscillator:
\begin{equation}
x_{0,{\rm zero\ point}} 
= \sqrt{\hbar \over \gamma m \omega} = \sqrt{\lambdabar_C \lambdabar \over
\gamma},
\label{eq11}
\end{equation}
where $\lambdabar = c/\omega$ is the laboratory (reduced) wavelength of
the transverse oscillation as measured along the beam axis.
In practical laboratory devices, we will have $\lambdabar \gg \gamma
\lambdabar_C$.  Hence,
the excitation of the transverse oscillations by fluctuations in the
radiation of the oscillating charge, as are described by the Hawking-Unruh
temperature, is negligible compared to the zero-point fluctuations of the
transverse oscillations.  In this sense, we can say along with Huang, Chen
and Ruth that the radiation does not excite the transverse oscillations, and
those oscillations will damp to the quantum-mechanical limit.  

In futuristic
devices, for which $\gamma > \lambdabar/\lambdabar_C$, \ie, when
\begin{equation}
\gamma > {mc^2 \over k \lambdabar_C},
\label{eq11a}
\end{equation}
quantum excitations of oscillations in a linear focusing channel would become
important.  When (\ref{eq11a}) holds, the transverse oscillations
would be relativistic even when their amplitude is only a Compton wavelength.
The strength of the transverse fields in the channel would 
then exceed the QED critical field strength (in the average rest frame),
\begin{equation}
E_{\rm crit} = {m^2c^3 \over e\hbar} = 1.6 \times 10^{16}\ \mbox{V/cm} 
= 3.3 \times 10^{13}\ \mbox{Gauss},
\label{eq6a}
\end{equation}
and the beam energy would be rapidly dissipated by pair creation.

Another way of viewing a practical linear focusing channel is that
its Hawking-Unruh excitation energy,
(\ref{eq8}), is small compared to the zero-point energy, $\hbar \omega^\star/2
= \gamma \hbar \omega/2$ of transverse oscillations.

The quantum-mechanical limit for transverse motion can, of course, also be 
deduced from the uncertainty principle:
\begin{equation}
\sigma_x \sigma_{p_x} \gsim \hbar,
\label{eq12}
\end{equation}
which leads to a minimum normalized emittance of
\begin{equation}
\epsilon_N = {\sigma_x \sigma_{p_x} \over mc} \approx \lambdabar_C,
\label{eq13}
\end{equation}
corresponding to geometric emittance of
\begin{equation}
\epsilon_x = {\epsilon_N \over \gamma \beta_z} \approx 
{\lambdabar_C \over \gamma}.
\label{eq14}
\end{equation}

\section{Semiclassical Analysis}

In a quantum analysis of a linear focusing channel, we found that the
transverse oscillations can damp to the limit set by the
uncertainty principle.  Hence, in a classical analysis we would expect
the damping to be able to proceed until the transverse amplitude was zero.

Indeed, a simple analysis confirms this.  Transform to the longitudinal rest
frame, in which the particle's motion is purely transverse.  The particle
has nonzero kinetic energy in this frame, but its average momentum is zero.
The radiation due to the transverse oscillation is reflection symmetric about 
the transverse plane in this frame, so the radiation carries away energy but 
not momentum.  With time, all of the energy would be radiated away, and the
particle would come to rest.   The transverse oscillations will have damped
to zero without affecting the longitudinal motion.

If we add the concept of photons to the preceding analysis, we can say that
the radiated photons carry away momentum along the direction of emission, but
the radiation pattern is symmetric, so the averaged radiated momentum is
zero.  Again, the radiation carries away energy, now in the form of photons.

Back in the lab frame, we view the photons as carrying away a small amount of
longitudinal
momentum on average, as a result of the Lorentz transformation of the
energy radiated in the longitudinal rest frame.  
This momentum, however, is only that part of the particle's longitudinal
momentum associated with its transverse oscillation; the longitudinal
velocity of the particle is unaffected.

On average, the photons carry away no transverse momentum in the lab frame, 
and the average momentum of the
radiated photons is therefore parallel to the beam axis in lab frame.
However, there is no need to argue that the momentum of individual radiated
photons is parallel to the beam axis, nor to imply that the matter of
the focusing channel absorbs transverse momentum in a manner than affects
the kinematics of the radiation process \cite{Huang2}.

\section{Comparison of a Linear Focusing Channel to a Wiggler}

A comparison with the behavior of particle beams in a wiggler is instructive.
Here the transverse confinement of the beam motion is provided by a series
of alternating transverse magnetic fields.  This has the notable effect that
even if a particle enters the wiggle parallel to the beam axis, transverse
oscillations will result whose amplitude is independent of the initial
transverse coordinate.  

In contrast, a particle that enters a linear focusing channel parallel to and 
along the axis undergoes no oscillation, no matter what is the particle's
longitudinal momentum.

We thereby see that radiation damping cannot reduce the
oscillations in a wiggler to zero unless the longitudinal momentum falls to 
zero also, since the wiggler continually re-excites
transverse oscillations for any particle with nonzero kinetic energy.

Another difference between a wiggler and a linear focusing channel can be seen 
by going to the longitudinal rest frame.  In the case of the wiggler, the
alternating magnetic fields in the laboratory transform to fields that are
very much like a plane wave propagating against the direction of the
laboratory motion of the beam.  The radiation induced by this effective
plane wave is not symmetric with respect to the transverse plane, but
results in a net kick of the particle into the backward direction.

Viewed in the lab frame, we find that along with the damping of their transverse
oscillations, the particles' longitudinal momenta are
significantly reduced.  To maintain the initial longitudinal momentum,
the beam must be reaccelerated. The momentum (and energy)
added back into the beam then increases the amplitude of the transverse
oscillations, and the damping cannot continue beyond some limit.

In contrast, in a linear focusing channel, the transverse damping proceeds
without significant reduction in the longitudinal momentum of the particle,
and the transverse oscillations can damp to the quantum limit without the
need of adding energy back into the beam.

%xxx

\section*{Acknowledgments}

I wish to thank Pisin Chen, Ron Ruth and Max Zolotorev for conversations
on radiation damping.
This work was supported in part by
DoE grant DE-FG02-91ER40671.

%\section*{Appendix}

\end{document}